\documentclass[draft]{agujournal2019}
\usepackage{url} 
\usepackage{lineno}
\usepackage{soul}
\usepackage{multicol}
\usepackage{multirow}

\usepackage{amssymb}
\usepackage{amsmath}
\usepackage[section]{placeins}

\usepackage{wrapfig}
\draftfalse

\journalname{JGR: Machine Learning and Computation}

\begin{document}

%
%

\title{A Multi-Model Ensemble System for the outer Heliosphere (MMESH): Solar Wind Conditions near Jupiter}

%
%




\authors{M. J. Rutala\affil{1}, C. M. Jackman\affil{1}, M. J. Owens\affil{2}, C. Tao\affil{3}, A. R. Fogg\affil{1}, S. A. Murray\affil{1}, L. Barnard\affil{2}}


\affiliation{1}{School of Cosmic Physics, DIAS Dunsink Observatory, Dublin Institute for Advanced Studies, Dublin, Ireland}
\affiliation{2}{Department of Meteorology, University of Reading, Earley Gate, PO Box 243, Reading RG6 6BB, UK}
\affiliation{3}{Space Environment Laboratory, National Institute of Information and Communications Technology (NICT), Koganei, Japan}




\correspondingauthor{M. J. Rutala}{mrutala@cp.dias.ie}




\begin{keypoints}
\item The performance of several existing solar wind propagation models at the orbit of Jupiter is measured for multiple spacecraft epochs.
\item A flexible system is developed to generate an ensemble of multiple propagation models so as to best leverage each input model's strengths.
\item Over the epoch tested, the multi-model ensemble outperforms individual input models by $7\%-110\%$ in forecasting the solar wind flow speed.
\end{keypoints}

%
%

%
%


\begin{abstract}
How the solar wind influences the magnetospheres of the outer planets is a fundamentally important question, but is difficult to answer in the absence of consistent, simultaneous monitoring of the upstream solar wind and the large-scale dynamics internal to the magnetosphere. To compensate for the relative lack of in-situ data, propagation models are often used to estimate the ambient solar wind conditions at the outer planets for comparison to remote observations or in-situ measurements. This introduces another complication: the propagation of near-Earth solar wind measurements introduces difficult-to-assess uncertainties. Here, we present the Multi-Model Ensemble System for the outer Heliosphere (MMESH) to begin to address these issues, along with the resultant multi-model ensemble (MME) of the solar wind conditions near Jupiter. MMESH accepts as input any number of solar wind models together with contemporaneous in-situ spacecraft data. From these, the system characterizes typical uncertainties in model timing, quantifies how these uncertainties vary under different conditions, attempts to correct for systematic biases in the input model timing, and composes a MME with uncertainties from the results. For the case of the Jupiter-MME presented here, three solar wind propagation models were compared to in-situ measurements from the near-Jupiter spacecraft \textit{Ulysses} and \textit{Juno} which span diverse geometries and phases of the solar cycle, amounting to more than 14,000 hours of data over 2.5 decades. The MME gives the most-probable near-Jupiter solar wind conditions for times within the tested epoch, outperforming the input models and returning quantified estimates of uncertainty.
\end{abstract}

\section*{Plain Language Summary}
The sun interacts with all the planets in the solar system through the solar wind, a stream of charged particles which blow outwards from the sun in all directions, carrying the interplanetary magnetic field with them. Both the magnetic field and particles interact with planetary magnetic fields with dramatic effects, including the aurora-- which shine not only on the Earth, but on gas giants of the outer solar system, like Jupiter, too. Characterizing the relationship between the solar wind and planetary magnetic fields is easiest with direct spacecraft measurements of both. Spacecraft between the Earth and Sun measure the solar wind, providing valuable context for understanding its interaction with the Earth. Unfortunately, there are no such permanent spacecraft near the other planets. Instead, models can be used to estimate the solar wind at these planets; however, these models can have significant, difficult-to-characterize uncertainties. Here we present the Multi-Model Ensemble System for the outer Heliosphere (MMESH), a framework designed to measure these uncertainties and attempt to correct for them by comparing multiple solar wind models to spacecraft measurements over a long time span. The final result here is an improved solar wind model, with estimated uncertainties, for Jupiter.

%
%

\section{Background}
The solar wind is a continuous stream of plasma emanating from the Sun in all directions which evolves as it travels through the heliosphere, interacting with every planetary magnetosphere in the solar system along the way. Near the Earth, the typical values of the solar wind flow speed $u_{mag, \oplus} (324-584 \textrm{ km/s})$, proton density $n_{\oplus} (2.2-12.7 \textrm{ cm}^{-3})$, dynamic (ram) pressure $p_{dyn, \oplus} (0.86-3.92 \textrm{ nPa})$, and interplanetary magnetic field (IMF) magnitude $B_{mag, \oplus} (3.1-9.7 \textrm{ nT})$ have all been statistically characterized by the expansive OMNI dataset \cite{King2005, Papitashvili2020}, with values here spanning the start of OMNI2 to the start of 2023 (1963/11/27 -- 2023/01/01) and characterizing $80\%$ ($10^{th}-90^{th}$ percentiles) of all measurements. The OMNI dataset is a composite of many near-Earth observations encompassing some $19$ total spacecraft over its full time domain, including most recently \textit{Wind} \cite{Lepping1995, Kasper2002, King2005, WilsonIII2021} and \textit{ACE} \cite{McComas1998, Smith1998, King2005}.

While fewer in-situ heliospheric data are available in the outer solar system, the average solar wind conditions have still been constrained by the various spacecraft to visit the outer planets, whether during planetary flyby or approach. At Jupiter, the most-visited of the outer planets, the solar wind has been characterized during the flybys of \textit{Pioneers 10} and \textit{11}, \textit{Voyagers 1} and \textit{2}, \textit{Ulysses}, \textit{Cassini}, and \textit{New Horizons} \cite<e.g.>{Slavin1985, Richardson1995, Hanlon2004_SolarWind, Ebert2010, Jackman2011, Ebert2014}. Compared to flybys, orbiter missions, including \textit{Galileo} and \textit{Juno} at Jupiter, generally provide fewer in-situ data: these missions have close-in orbits to best study the planet itself, thus setting them deep inside the planet's magnetosphere and shielding them from the solar wind. As a result, they only sample the wind during the planetary approach phase prior to orbital insertion and occasional excursions into the solar wind near apoapsis. 
The \textit{Ulysses} spacecraft, which used near-Jupiter orbital maneuvers to enter its distant, polar trajectory, gives the best single-spacecraft characterization of the average near-Jupiter solar wind owing to its $18$-year lifetime: $80\%$ of \textit{Ulysses} measurements span $380-520\textrm{ km/s}$ in solar wind flow speed $u_{mag}$, $0.05-0.55\textrm{ cm}^{-3}$ in plasma density $n$, $0.02-0.20\textrm{ nPa}$ in dynamic pressure $p_{dyn}$, and $0.22-1.5\textrm{nT}$ in IMF magnitude $B_{IMF}$ \cite{Ebert2014}. Despite the large number of measurements, these distributions represent a biased sample fo the near-Jupiter solar wind due to the polar orbit of the \textit{Ulysses} spacecraft; \textit{Ulysses} samples the solar wind in the ecliptic plane periodically, and these numbers were drawn from two non-consecutive spans at different phases of the solar cycle-- one with a slower, cooler, and denser average solar wind than the other \cite{Ebert2014}.

The highly dynamic nature of the solar wind is not captured by these average values. Singular events, such as the eruption of coronal mass ejections (CMEs), and their propagation through the heliosphere as interplanetary coronal mass ejections (ICMEs), are a major source of short timescale variation in the measured solar wind \cite[and references therein]{Palmerio2021}. In terms of the quantities already discussed, interplanetary coronal mass ejections (ICMEs) show expansion, which manifests in measurements as an increase in $u_{mag}$ at the leading edge and a decrease at the trailing edge, large drops in $n$, and an enhancement in $B_{IMF}$ magnitude but decrease in $B_{IMF}$ variance \cite{Zurbuchen2006, Owens2018_ICME}. Beyond these events, the ambient solar wind is dynamic due to the presence of two different streaming plasma populations originating in different regions of the solar corona: a comparatively fast, hot, and tenuous stream and a comparatively slow, cool, and dense stream \cite{Crooker1999}. These streams are essentially bimodal during solar minimum, with fast streams originating at high heliolatitude and slow streams originating nearer the solar equator \cite{McComas1998, McComas2000}; during solar maximum, these streams are markedly less ordered \cite{McComas2003}. From solar cycle to solar cycle, the bulk parameters of the fast stream in particular can change dramatically \cite{McComas2008, Ebert2009, McComas2013, Ebert2014}. As different regions of the sun rotate underneath, corotating interacting regions (CIRs) are formed where a fast flow catches up to a slow flow; this process is common throughout the heliosphere \cite{Richardson2018} and drives significant interactions with planetary magnetospheres, including at the Earth \cite{Crooker1999, Gosling1999, Tsurutani2006, Borovsky2010}, at Jupiter \cite{McComas2003, Hanlon2004_SolarWind, Ebert2014}, and beyond \cite{Jackman2004}.

In-situ-data-driven statistical studies of the time variable solar wind at specific locations within the outer heliosphere (e.g. at Jupiter) are hampered by the limited temporal coverage of visiting spacecraft; there is no continuous composite dataset like OMNI for any outer planet. Such statistical studies often instead have solar wind data supplemented by solar wind propagation models, which attempt to reproduce the time-varying solar wind at one location from measurements at another location at which the solar wind is known. Many of these models have been employed in the outer heliosphere, including, but not limited to, the model of \citeA{Tao2005} (``Tao+'', hereafter), ENLIL \cite{Odstrcil2003}, mSWiM \cite{Zieger2008}, HUXt \cite{Barnard2022, Owens2020}, and MSWIM2D \cite{Keebler2022}. These models all differ in their dimensionality, the simplifications made to the magnetohydrodynamics (MHD) equations underlying them, and the source of the input solar wind conditions used to initialize the model. By virtue of modelling solar wind conditions for times and locations where no in-situ spacecraft measurements are available, the outputs of these models cannot be directly compared to data in typical usage scenarios. Generally, solar wind propagation models are instead compared to in-situ spacecraft measurements at times and locations where they are available in order to approximate the model errors-- generally, shock arrival time (or ``timing'') errors-- prior to being used to supplement the data \cite{Tao2005, Zieger2008, Keebler2022}. Measured timing uncertainties can be as high as $\pm4$ days and often trends with other physical parameters of the system, such as with Target-Sun-Observer (TSO) angle \cite{Tao2005, Zieger2008, Keebler2022} or with phase of the solar cycle \cite{Zieger2008}.

These resulting time-varying timing uncertainties introduce a challenge in interpreting the results of these models and performing statistical analyses, particularly because the characterizations of timing uncertainty in each propagation model are often not measured by the same methods, and thus are not directly comparable to one another.
Timing uncertainties can be measured by manually identifying shocks and shock-like structures in both modeled and measured solar wind time series and comparing their occurrence times {\cite<e.g.>{Tao2005}} or by offsetting one time series relative to the other and maximizing the resulting prediction efficiency, or Pearson correlation coefficient {\cite<e.g.>{Zieger2008}}. Measuring uncertainties with the latter method implies that a single timing uncertainty characterizes the model over the full time period inspected. An alternative to this is to employ dynamic time warping to explicitly allow for time-varying timing uncertainties {\cite<e.g.>{Samara2022}}. If these model uncertainties were quantified in a cross-model-consistent manner, the time-varying uncertainties could be accounted for and partially mitigated. For instance, as propagation model output uncertainties are known to trend with physical quantities, each individual model's outputs could be de-trended with sufficient characterization of the uncertainties. Alternatively, a multi-model ensemble (MME) could be composed by cross-comparison of the models in order to mitigate uncertainties.
An MME is, in essence, a weighted average of different model outputs \cite{Murray2018}; the weighting scheme can be adjusted based on metrics of the performance (or ``skill") of the models during intervals where in-situ data are available \cite{Murray2018, Elvidge2023}. Ideally, fully-independent models would be used in an MME, so that they would be expected to have independent random errors which would thus tend to cancel, rather than add \cite{Hagedorn2005, Riley2018}. If all input models capture the same physics, outperform one another in different parameter spaces, and have independent errors, a MME of these models should describe the underlying physical system more accurately than any individual input.

Here we present the Multi-Model Ensemble System for the outer Heliosphere (MMESH): a framework to quantify and mitigate timing uncertainties in solar wind propagation models and produce a single prediction by combining all of these approaches. This system allows for the automatic  quantification of model timing uncertainties, trending of timing uncertainties with physically relevant parameters, de-trending of the original model timing, and combination of distinct models into a single MME. MMESH is designed to flexibly compare any combination of input solar wind propagation models and contemporaneous in-situ data in order to create an MME. To demonstrate this concretely, here we construct an MME of the solar wind conditions at Jupiter during the \textit{Juno} era.

Thus prior to discussing MMESH itself, we first discuss the in-situ spacecraft datasets to be used for comparison (Section \ref{sec:Inputs:Data}) and give some introduction to the specific solar wind propagation models considered here (Section \ref{sec:Inputs:Models}). We then introduce the MMESH framework in Section \ref{sec:MMESH}, beginning with a description of the statistical techniques and tools used to compare models, including the MME, to contemporaneous data and measure their performance (Section \ref{sec:MMESH:PerformanceMetrics}. In Section \ref{sec:MMESH:Performance} we discuss the methods available to characterize the model timing uncertainties relative to the in-situ time series: constant time offsetting (Section \ref{sec:MMESH:Performance:Constant}) and dynamic time warping (DTW, Section \ref{sec:MMESH:Performance:DTW}). We then proceed to describe how trends in the empirical timing uncertainties are characterized and estimated for epochs without contemporaneous in-situ data (Section \ref{sec:MMESH:Prediction}) before discussing the composition (Section \ref{sec:MMESH:MME}) and performance (Section \ref{sec:MMESH:MMEPerformance}) of the multi-epoch MME composed of the de-trended models. Having described MMESH, we then present the MME of the solar wind conditions at Jupiter for the first $7$ years of the \textit{Juno} mission, spanning $2016/07/04 - 2023/07/04$, for use in future statistical analyses (Section \ref{sec:JunoEpochResults}), prior to concluding.

\section{Inputs} \label{sec:Inputs}

\subsection{Solar Wind Data}\label{sec:Inputs:Data}

The present aim for the MME framework discussed here is to find the most accurate combination of solar wind models in the near-Jupiter region of the outer heliosphere. As such, limiting the data included for comparison to the input and ensemble models to that which is representative of conditions at Jupiter is essential. Including too large a range of radial or helio-latitudinal in-situ measurements risks including different regimes of solar wind properties which the models are not, and should not be, expected to reproduce. This is particularly an issue in choosing a useful range of heliolatitude-- too narrow a range and the amount of data available shrinks, but too large a range and the faster solar wind flows at higher heliolatitudes are included erroneously. This issue primarily relates to data acquired by the \textit{Ulysses} spacecraft, which is a solar polar orbiter. Previous \textit{Ulysses} measurements show that, during solar minimum when the latitudinal structure of the solar wind is well-ordered, the equatorial slow solar wind zone may extend to $\pm20^\circ-\pm30^\circ$ about the solar equator \cite{McComas2003}. \citeA{Ebert2014} further restricts this range in surveying near-Jupiter solar wind conditions measured with \textit{Ulysses} and selects for data $\pm10^\circ$ about the solar equator. Measurement-driven models of solar wind variability in the near-Earth environment, which show that solar wind properties are highly localized during solar minimum, varying significantly over ${\sim}2^{\circ}$ heliolatitude, underscore the importance of carefully selecting a heliolatitude range to consider \cite{Owens2020_heliolatitude}.

\begin{table}[!t]
\caption{In-situ measurements of solar wind parameters near Jupiter's orbit.}
\centering
\begin{tabular}{r | l | c | c | c }
\hline
Mission & Coverage & Range & Heliolatitude & Measurements \\
(Epoch) &  [yyyy/mm/dd] & [AU] & [$\deg$] & [hr] \\
\hline
   \textit{Ulysses} (01) & 1991/12/08 -- 1992/02/02  & $4.90-5.41$ &  $-6.10$ -- $+6.10$ & 1,344 \\
   (02) & 1997/08/14 -- 1998/04/16 & $4.90-5.41$ & $-6.10$ -- $+6.10$ & 5,878 \\
   (03) & 2003/10/24 -- 2004/06/22 & $4.90-5.41$ & $-6.10$ -- $+6.10$ & 5,801 \\
   \textit{Juno} \hspace{5mm} & 2016/05/15 -- 2016/06/29 & $5.27-5.44$ & $-5.76$ -- $-5.23$ & 1,080 \\
\hline
\end{tabular}
\label{tab:Spacecraft}
\end{table}

Here, the near-Jupiter outer heliosphere is defined as the region of the heliosphere spanning $4.9 \textrm{ AU} < r < 5.5 \textrm{ AU}$ for spherical distance from the Sun $r$ and $-6.1^\circ \leq \theta \leq 6.1^\circ$ for heliolatitude $\theta$. Jupiter's perihelion and aphelion ($5.04$ and $5.37$ AU, respectively) fit entirely within this range, which includes padding of $\sim0.15 \textrm{ AU}$, or approximately $50\%$, on either end to increase the number of observations included. The heliolatitude range selected represents the maximal range of Jupiter's location in heliolatitude without any padding in order to avoid unrealistic sampling of the high latitude fast solar wind flows.

\begin{figure}[!ht]
    \centering
    \includegraphics[width=\textwidth]{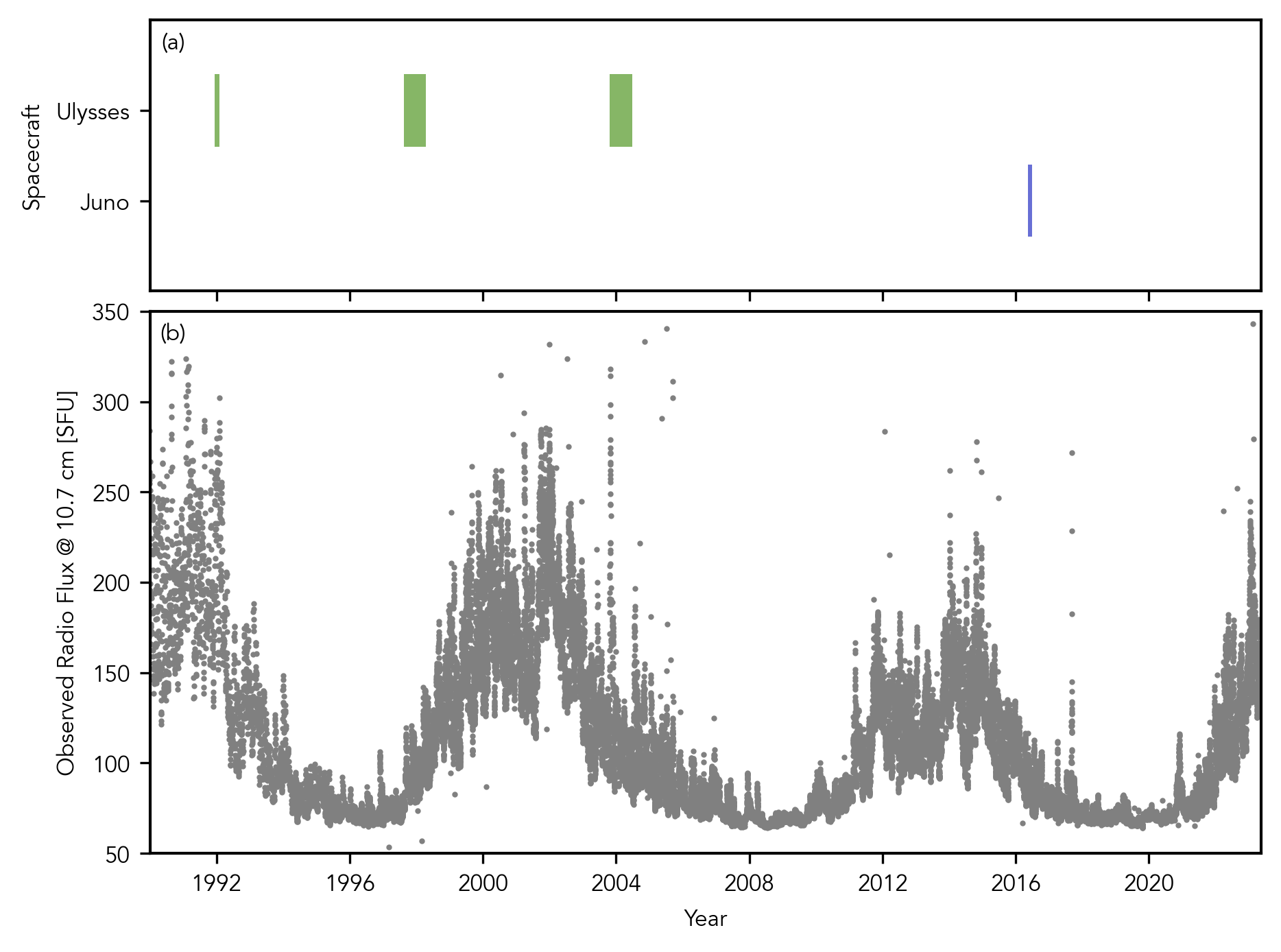}
    \caption{The (a) spans during which each spacecraft used in this analysis (\textit{Ulysses}, in green, and \textit{Juno}, in blue) was measuring the near-Jupiter solar wind compared with the (b) solar F10.7 cm radio flux, a proxy for the phase of the solar cycle, over the period 1990-2023. Spacecraft coverage spans the ascending and descending phases of the solar cycle, but largely excludes solar minimum and solar maximum. These spacecraft have been selected for the cadence of their plasma and magnetic field measurements, which are generally hourly or better.}
    \label{fig:SpacecraftSolarCycle}
\end{figure}

\begin{figure}[!ht]
    \centering
    \includegraphics[width=\textwidth]{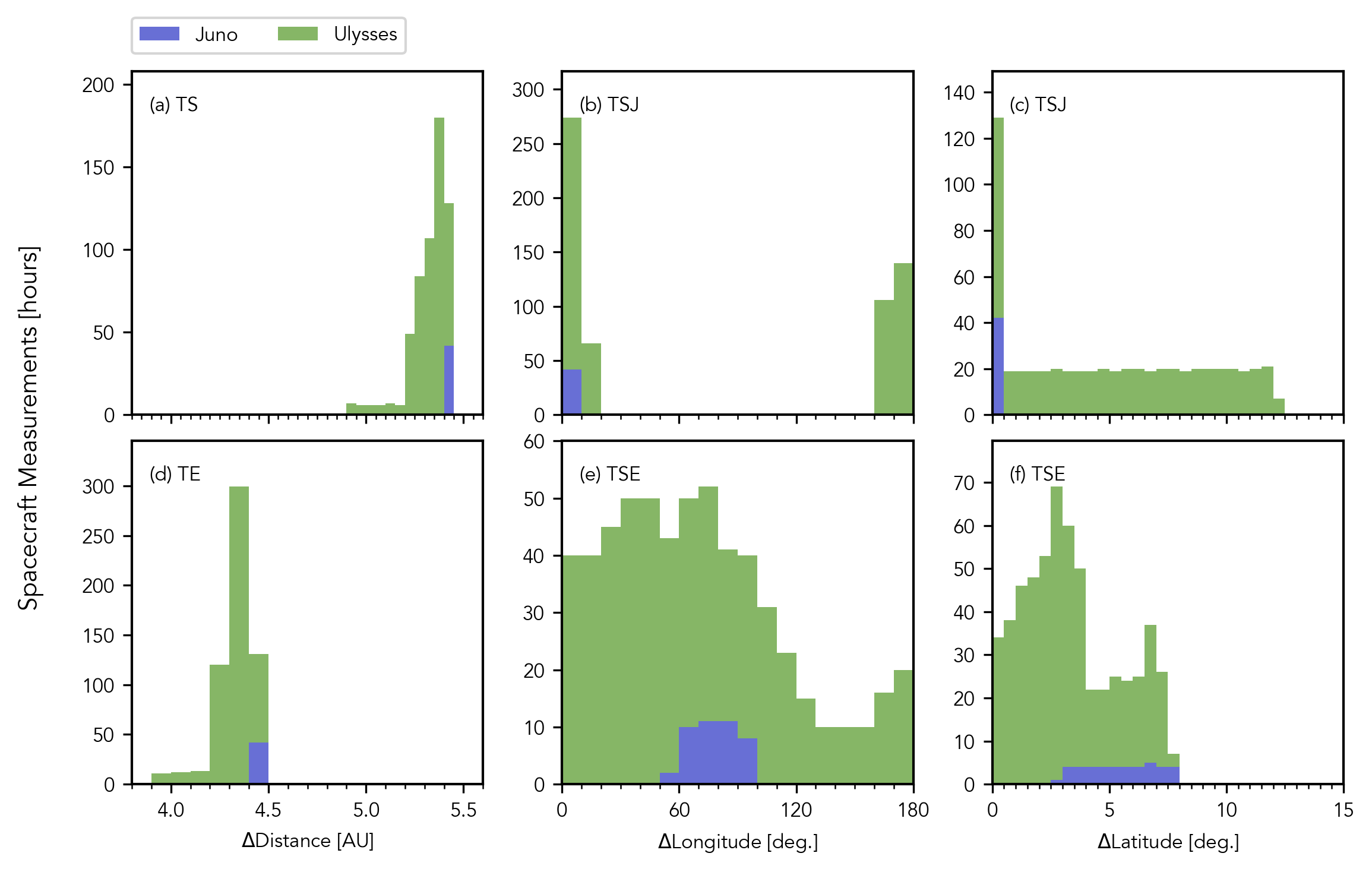}
    \caption{Histograms showing the spatial coverage of all the spacecraft (or `targets') used here, including the (a) Target-Sun (TS) distance, (b) Target-Sun-Jupiter (TSJ) longitude angle, (c) TSJ latitude angle, (d) Target-Earth (TE) distance, (e) Target-Sun-Earth (TSE) longitude angle, and (f) TSE latitude angle. The angles are measured in the Sun's inertial reference frame, such that longitude measures distance along the solar equator and latitude measures perpendicular distances along the sphere of the Sun. The majority of spacecraft measurements occur very near Jupiter, with minimal separation in TSJ longitude or latitude angles. The unique coverage of the polar-orbiting \textit{Ulysses} spacecraft stands out, and provides even coverage across TSJ, and to a lesser extent TSE, latitudes. Taking all spacecraft into consideration, the spatial coverage relative to the Earth's location is fairly even.}
    \label{fig:Trajectories}
\end{figure}

Several spacecraft have transited this region, including \textit{Pioneers 10} and \textit{11}, \textit{Voyagers 1} and \textit{2}, \textit{Ulysses}, \textit{Galileo}, \textit{Cassini}, \textit{New Horizons} and \textit{Juno}. Here, just data from just the \textit{Ulysses} and \textit{Juno} missions are used; the remaining spacecraft are not used in this analysis either due to being discontinuous at temporal resolutions of 1 hour (\textit{Galileo}, \textit{Cassini}, and \textit{New Horizons}) or due to a lack of coverage in all or some of the models to be discussed in Section {\ref{sec:Inputs:Models}} (\textit{Pioneers 10} and \textit{11} and \textit{Voyagers 1} and \textit{2}). A brief overview of the used spacecraft trajectories and data is included in Table \ref{tab:Spacecraft}. The durations of the visits of these spacecraft to the near-Jupiter outer heliosphere is illustrated in Figure \ref{fig:SpacecraftSolarCycle} relative to the solar cycle, as measured by F10.7 radio flux derived from observations at the Dominion Radio Astrophysical Observatory (DRAO) and adjusted to account for variations in the Earth's distance from the Sun. While the majority of these spacecraft passed near Jupiter, the \textit{Ulysses} spacecraft, as a polar orbiter, transits through the near-Jovian outer heliosphere away from the planet itself after its initial Jupiter flyby. The relevant orbital components for all the spacecraft in Table \ref{tab:Spacecraft} are shown in Figure \ref{fig:Trajectories}, which highlights the rarity of near-Jupiter outer heliosphere measurements made far from Jupiter itself and the comparative evenness of coverage in Target-Sun-Earth angle. 

All of the spacecraft referenced in Table \ref{tab:Spacecraft} have both magnetometers and plasma instruments, and thus provide sampling of the interplanetary magnetic field (IMF) $B_{IMF}$, the solar wind ion number density $n$, and the magnitude of the solar wind flow speed $u_{mag}$, which is itself dominated by the radial component of the outward flow of the solar wind. As the proton density $n_{p}$ is measured in all cases and protons are the dominant ion component of the solar wind \cite<e.g.>{Ebert2014}, the total density of the solar wind is approximately equal to the proton density ($n \approx n_p$) and is assumed to be exactly equal in calculating the solar wind dynamic pressure $p_{dyn} = m_p n u_{mag}^2$, where $m_p$ is the proton mass. Detailed descriptions of these instruments, including their heritages, limitations, and data products, are discussed in their respective instrument papers \cite{Balogh1992, Bame1992, Connerney2017, McComas2017}.

\subsection{Solar Wind Models}\label{sec:Inputs:Models}

While several solar wind propagation models for the outer heliosphere are available, three were chosen for detailed study and inclusion in the MME: the Tao+ \cite{Tao2005}, ENLIL \cite{Odstrcil2003}, and HUXt \cite{Owens2020, Barnard2022} models. These models in particular are ideal for inclusion in a MME due to their differing input parameters, dimensionality, and approaches to propagating the solar wind beyond the Earth, as is summarized in Table \ref{tab:models} and will be discussed further here.

\begin{table}[!ht]
\caption{Descriptive parameters of solar wind propagation models as used in this study.}
\centering
\begin{tabular}{l l l l l l}
\hline
Model & Type & Inner Boundary$^a$ & MHD Terms$^{b}$ & Input$^a$ & Output$^{c}$ \\
 & & [AU] & (P, L, G, C) & type (source) & $(n, u_{mag}, p_{dyn}, B_{IMF})$ \\
\hline
    ENLIL & 3D MHD & ${\sim}0.1$ & (P, L, G) & remote (WSA) & $(n, u_{mag}, p_{dyn}, B_{IMF})$ \\ 
    HUXt & 1D HD & ${\sim}1$ & -- & in-situ (OMNI) & $(u_{mag})$ \\ 
    Tao+ & 1D MHD & ${\sim}1$ & (P, L, G) & in-situ (OMNI) & $(n, u_{mag}, p_{dyn}, B_{IMF})$ \\ 
\hline
    \multicolumn{6}{l}{$^a$ Inner boundaries and input types are reported for the versions of the models used here. The} \\
    \multicolumn{6}{l}{models are not necessarily limited to these inner boundaries and input types only, as described} \\
    \multicolumn{6}{l}{in the text.} \\
    \multicolumn{6}{l}{$^b$ The (P)ressure, (L)orentz, (G)ravitational, and (C)ollisional terms of the governing MHD} \\
    \multicolumn{6}{l}{momentum equation (Eqn. \ref{eqn:momentum_full}).} \\
    \multicolumn{6}{l}{$^c$ Components of the solar wind: plasma density ($n$), plasma flow speed ($u_{mag}$), plasma} \\
    \multicolumn{6}{l}{dyamic pressure ($p_{dyn}$), or IMF ($B_{IMF}$).} \\
\label{tab:models}
\end{tabular}
\end{table}

Fundamentally, most models propagate solar wind conditions outwards by solving the system of equations which constitute MHD, these being: the mass continuity equation, the momentum equation, the equation of state, and several physical laws necessary to close the system (Faraday's, Ohm's, and Amp\`ere's laws). Propagation models differ primarily in their treatment of the momentum equation. For a single-species plasma composed of protons, this is:
\begin{equation}
    \frac{\partial (m_p n \vec{u})}{\partial t} 
    + \rho(\vec{u} \cdot \nabla) \vec{u}
    = - \underbrace{\nabla p}_{pressure} 
    + \underbrace{\vec{j} \times \vec{B}}_{Lorentz}
    - \underbrace{\frac{GM_{\odot}\rho}{r^2} \hat{r}}_{gravity} + \underbrace{\nu \nabla^2 \vec{u}}_{collision}
    \label{eqn:momentum_full}
\end{equation}
where $m_p$ is the proton mass, $n$ is the plasma number density, $\vec{u}$ is the plasma flow velocity, $p$ is the total plasma pressure, $\vec{j}$ is the plasma current density, $\vec{B}$ is the ambient magnetic field, $G$ is the gravitational constant, $M_{\odot}$ is a solar mass, $r$ is the radial distance in a heliocentric spherical frame with the $\hat{r}$ direction pointing radially outward, and $\nu$ is a collisional frequency. In Equation \ref{eqn:momentum_full}, the right-hand-side terms are labelled corresponding to the physical forces they represent, these being the (gradient) pressure, Lorentz, gravitational, and collisional forces, respectively. As summarized in Table \ref{tab:models}, solar wind propagation models differ in which terms of the momentum equation they assume are insignificant in the solar wind. Most propagation models, including all those discussed here, do not consider collisional forces within the solar wind plasma. Both ENLIL and Tao+ keep all the remaining terms shown in Equation \ref{eqn:momentum_full} \cite{Tao2005, Odstrcil2003}. HUXt assumes that all forces are negligible compared to the magnitude of the left-hand-side momentum terms in Equation \ref{eqn:momentum_full}, and thus does not consider any force terms \cite{Owens2020}.

The variables propagated by each model are directly related to the force terms that they consider in Equation \ref{eqn:momentum_full}, and are listed in Table \ref{tab:models} for the three models discussed here. The dimensionality of each model changes which components of the vector terms in Equation {\ref{eqn:momentum_full}} can be propagated; for cross-model consistency, we therefore compare solar wind parameter magnitudes rather than vector components, where each magnitude is calculated as the root-sum-square of available components. The solar wind flow speed $u_{mag}$ is thus available from all three propagation models considered here. From HUXt, none of the solar wind density $n$, temperature $T$, or IMF strength $B_{IMF}$ are propagated, as these variables are eliminated from the version of the momentum equation used. These parameters-- density $n$, temperature $T$, and IMF strength $B_{IMF}$ of the propagated solar wind-- are available from both ENLIL and Tao+.

Each of these models has an inner boundary at which the conditions of the solar wind are input and continuously updated over the course of the model run. The location of this inner boundary and the sources from which the input solar wind conditions are drawn vary between models and are summarized in Table \ref{tab:models}. ENLIL takes as input a 3-dimensional description of the solar corona and near-sun environment, here supplied by the Wang-Sheeley-Arge (WSA) model {\cite{Arge2000}} which itself takes remote observations of the Sun as input. For this study, solar magnetograms from the Kitt Peak Observatory are used, with gaps in observations filled in by those from the Mount Wilson Observatory. Together, these two observatories provide magnetograms up to Carrington Rotation 2196 (2017/11/06); after this point, ENLIL outputs at Jupiter are only available using Global Oscillations Network Group (GONG) observations. These GONG observations are not used here, due to ongoing issues in coupling them to WSA-ENLIL. This sort of boundary is unique amongst the models considered here: HUXt and Tao+ instead take in-situ spacecraft measurements, or proxies thereof, as inputs. In this study, both models take OMNI measurements at ${\sim}1$ AU as inputs, although they both have the functionality to be run at any other location in the solar system, provided there are sufficient in-situ solar wind data available {\cite<e.g.>{Sanchez-Diaz2016, Barnard2022}}. The solar wind conditions used as model input, and how well they reflect physical conditions, are the single largest factor in determining the accuracy of the propagated solar wind \cite{Riley2018}, and as such including a variety of inputs is beneficial to the final MME.

The input solar wind conditions used here are assumed to be sampled from the background solar wind. This means that coronal mass ejections (CMEs) sampled at the model inner boundary are not propagated using the standard cone model \cite{Zhao2002, Xie2004} but are instead interpreted as fast solar wind flows; rather than propagating CMEs as radially-expanding regions of constant angular size, they are treated by the same fluid description used by each model to describe the rest of the solar wind flow. This introduces an intrinsic error into the background solar wind parameters in all of the models. Future studies could mitigate this additional source of error by subtracting CMEs from the input data prior to propagation, then simultaneously propagating the quiescent solar wind and the CME using the cone model, but such an involved change to the modeling is ultimately beyond the current scope of this project.

These three models each run at different spatial and temporal resolutions which are directly related to their dimensionality and domains within the heliosphere, and which directly impact the small-scale shape of their output propagated solar wind estimates. ENLIL covers three spatial dimensions, spanning $0.1-10$ AU radially at $0.02$ AU resolution, $360^{\circ}$ in longitude at $2^{\circ}$ resolution, and $\pm60^{\circ}$ in latitude at $2^{\circ}$ resolution, with a temporal resolution of $1$ hour. HUXt is physically a one-dimensional radial model, but in practice here it is run in its two-dimensional form in order to more easily sample the model at the spacecraft position. Functionally, the two-dimensional form of HUXt is a series of independent one-dimensional models spanning $1-6$ AU radially with a resolution of $0.007$ AU, $360^{\circ}$ in longitude at ${\sim}2.8^{\circ}$ resolution, and an intrinsic temporal resolution of $17.4$ minutes in the version of the model used here. Tao+ spatial dimension, ranging from $1-8$ AU at a resolution of $1/300$ AU, with an intrinsic temporal resolution of $10$ s. The outputs of both HUXt and Tao+ have been downsampled to a resolution of $1$ hour to better match the spacecraft data and other models for use in this study. When sampling model parameters at the location of a target spacecraft or planet, the parameters are interpolated to the target's position. In all of these models, the fixed angular widths used for longitudinal bins mean that the area (or volume, for three-dimensional ENLIL outputs) of each model bin increases with increasing radial distance, thus increasing the distance between interpolation points and adding uncertainty in the modeled solar wind behavior. This represents one source of uncertainty which needs to be characterized to accurately estimate the effects of the solar wind on outer heliosphere targets.

Figures \ref{fig:ModelVelocities}a-c show the model-propagated solar wind flow speed $u_{mag}$ during the \textit{Juno} cruise towards Jupiter compared with contemporaneous JADE in-situ measurements from \citeA{Wilson2018} for each of the models detailed here. While these models are all able to propagate solar wind conditions during the other spacecraft epochs shown in Table \ref{tab:Spacecraft}, and both ENLIL and Tao+ are able to propagate parameters other than $u_{mag}$, here we have chosen to show just a single-spacecraft and single-parameter comparison for illustrative purposes. There is a general agreement between each model and the data in form and magnitude, but significant deviations in the arrival time of large-scale shocks and smaller-scale increases in flow speed between model and data, as can be seen near day of year 141 in Figures {\ref{fig:ModelVelocities}}a-c. These temporal lags, which represent single measurements of the full distribution of model timing uncertainties, appear to be of the same sign for Tao+ and HUXt but are substantially different for ENLIL. Characterizing these differences in arrival time is critically important to understanding the accuracy of these models in propagating the solar wind, and will be further explored here.

\section{Description of MMESH}\label{sec:MMESH}

The disagreements between the propagation models and in-situ data in both the modeled arrival time and magnitude, as illustrated in Figure \ref{fig:ModelVelocities}, makes the need for careful consideration of uncertainties and new statistical approaches in solar wind propagation modeling evident. MMESH has been designed as a framework to tackle these issues. After briefly introducing the statistical metrics used in quantifying model performance (Section \ref{sec:MMESH:PerformanceMetrics}), the MMESH framework will be described. This system allows any number of solar wind propagation models to be compared to simultaneous in-situ data; from this comparison, timing uncertainties are characterized either as a constant value over the full duration of each model (i.e. as a bias, as explored in Section \ref{sec:MMESH:Performance:Constant}) or as a dynamic value (Section \ref{sec:MMESH:Performance:DTW}). MMESH fundamentally supports a multi-epoch analysis, in which the same timing uncertainties are quantified over multiple spacecraft epochs, each with one set of in-situ data and multiple models, in order to better characterize the model timing uncertainties, including any timing biases (Section \ref{sec:MMESH:Prediction}). From this characterization, the solar wind propagation models can then have any identified biases in timing removed before being assemble into an MME (Section \ref{sec:MMESH:MME}).

\subsection{Performance Metrics}\label{sec:MMESH:PerformanceMetrics}

The correlation coefficient $r$, as a robust measure of model goodness-of-fit, is a good metric to be maximized in optimizing the alignment of solar wind model to data, as will be discussed in Section {\ref{sec:MMESH:Performance:Constant}}. For simple methods of aligning the model and data, the correlation coefficient $r$ is sufficient alone as a metric. More complex methods of alignment, such as discussed in Section {\ref{sec:MMESH:Performance:DTW}}, are better optimized while considering some penalty against increasing complexity, in order to maintain physical realism and interpretability. In this case, a statistic determined by both the correlation coefficient and some measure of the width of the distribution of timing uncertainties (as will be discussed Section {\ref{sec:MMESH:Performance:DTW}} and in Figure {\ref{fig:DTW_Widths}}) is preferred for optimization. Such a statistic is less likely to reach its maximum value when a large range of timining uncertainties are predicted, thus preventing unphysical alignment of a model with, for instance, a shock-like structure from a previous Carrington rotation. Within MMESH, we define $\sigma_T$ to be the half-width containing $34\%$ of the distribution of timing offsets, such that it would reduce to one standard deviation in a normal distribution. The optimization metric for these cases is then defined as $r + (1 - \sigma_T/\Delta T)$, where $\Delta T$ represents half the largest allowed magnitude of a timing uncertainty, such that the statistic varies between $0-2$, with the former corresponding to the worst performance and the latter corresponding to the best.

\begin{figure}
    \centering
    \includegraphics[width=\textwidth]{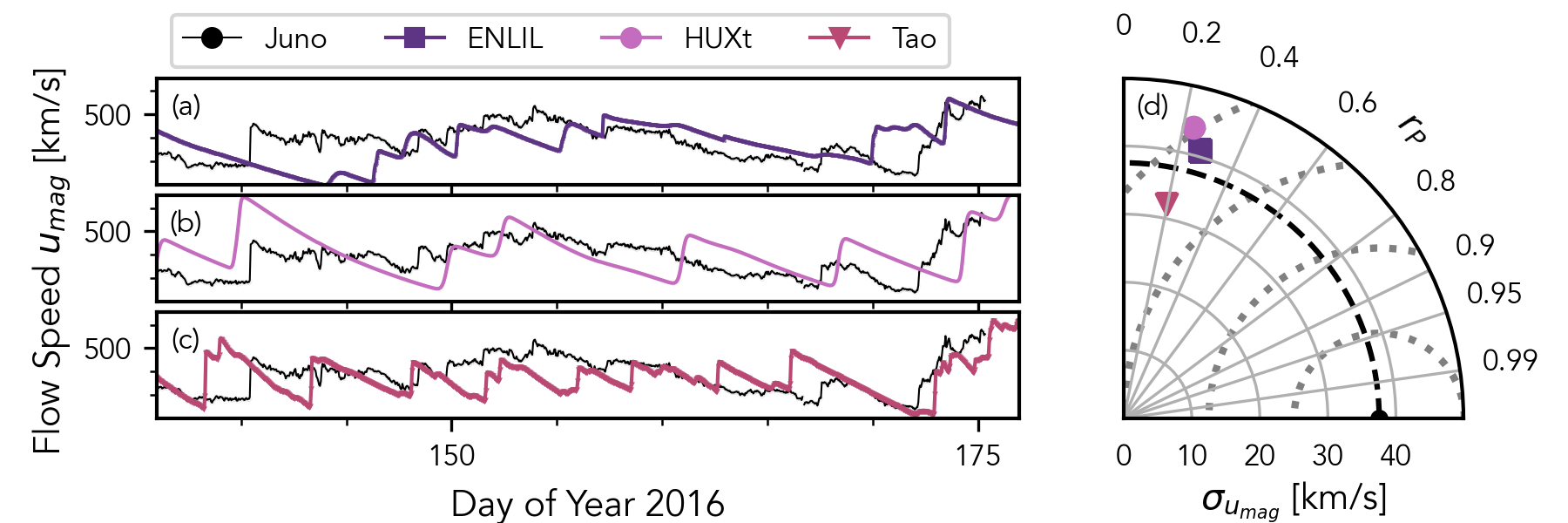}
    \caption{Measured solar wind flow speed $u_{mag}$ from \textit{Juno} JADE moments spanning 2016/05/15 - 2016/06/29 \cite{Wilson2018} with the same from the (a) ENLIL, (b) HUXt, and (c) Tao+ models, as labeled, with (d) a Taylor diagram illustrating the performance of each model relative to the data, as discussed in Section \ref{sec:MMESH:PerformanceMetrics} The flow speed is referenced as the root-mean-square of all velocity components, where components are available. Temporal lags in the timing of the modeled solar wind flow speed $u_{mag}$ are apparent in all models, and are made evident by the Taylor diagram.}
    \label{fig:ModelVelocities}
\end{figure}

Neither the correlation coefficient $r$ nor the statistic $r/\sigma_T$ as single numbers fully characterize how closely a model matches data on different scales. Combining the correlation coefficient $r$ with the overall standard deviation of both the time series and the model residuals forms the basis for a more complete multi-scale comparison of model and data summarized by the Taylor diagram \cite{Taylor2001}, illustrated in Figure \ref{fig:ModelVelocities}. This type of plot relates the standard deviation of the modeled time series, the correlation coefficient of the modeled time series relative to the measured time series, and the centered root-mean-square difference between the modeled and measured time series to one another by analogy with the law of cosines, allowing all three quantities to be displayed as a single point on the diagram. This is particularly useful for comparing the performance of different models to one another on the same axes. All-around better models-- those with high correlation coefficients, small residuals when compared with the data, and similar intrinsic variances-- appear graphically closer to the point representing the data time series along the x-axis.

\subsection{Characterization of Propagation Model Performance}\label{sec:MMESH:Performance}

The arrival time of shocks is of particular interest in statistical studies both at Jupiter and elsewhere in the outer heliosphere; the arrival of a shock is expected to compress the magnetosphere, directly impacting plasma and magentic flux transport and auroral activity \cite{Southwood2001, Cowley2003_me, Vogt2019, Nichols2019, Kita2019}. While individual models typically quote some uncertainty in modeled arrival times \cite{Tao2005, Zieger2008, Owens2020}, these uncertainties are often characterized relative to different standards and using different methods, making cross-model comparisons difficult. 

To allow direct comparisons of outer heliosphere solar wind models, independent quantification of modeled arrival time uncertainty can be performed with MMESH, as is common for near-Earth solar wind modeling \cite{Gressl2014, Riley2018}. The goal in quantifying the arrival time uncertainty is twofold. First, understanding the error intrinsic to each model is necessary to give context to its forecasts; second, characterizing these errors can give clues as to which aspects of the solar wind system an individual model may not be capturing sufficiently. For both of these reasons, here we explore two methods available in MMESH of quantifying the arrival time uncertainties in the previously discussed models. These comparisons and uncertainty characterization are performed identically for every combination of spacecraft and model previously discussed.
To keep both illustrations and discussion informative and uncluttered, here we will focus on comparisons to the \textit{Juno} in-situ measurements of the solar wind flow speed $u_{mag}$. The flow speed is a key descriptor of the overall solar wind behavior, particularly in regards to CIRs, and an indicator of solar wind-magnetosphere coupling, making it a highly relevant variable to consider in describing model timing uncertainties.

\subsubsection{Constant Time Offsetting}\label{sec:MMESH:Performance:Constant}

A simple metric to characterize the performance of a propagation model is to calculate the prediction efficiency, or correlation coefficient, between the propagated time series and an in-situ measurement of the same quantity \cite[e.g.]{Zieger2008, Keebler2022}. This offers a straightforward method to determine systematic, spacecraft-epoch-wide propagation model errors in the arrival time of shocks and other solar wind structures \cite<e.g.>{Fogg2023}. The time span covered by the model can be shifted off that of the measured data by an offset time $\Delta t$ both forward (i.e. later) and backward (i.e. earlier) in time, then the correlation coefficient between this offset model propagated time series and the in-situ measurements can be calculated and compared to the original. 

Performing this $2n+1$ times for temporal offsets spanning the values $[-n, -n+\Delta t, ..., n-\Delta t, n]$ for a realistic maximum offset time of $n\approx4$ days \cite{Tao2005, Zieger2008} yields the correlation coefficient as a function of constant temporal offsets, $r(\Delta t)$, with positive temporal offsets indicating that the un-offset model leads the data and negative offsets indicating that the un-offset model lags the data. Maximizing the correlation coefficient $r(\Delta t)$ thus gives a constant temporal offset which best aligns the propagation model with the measured time series. Equivalently, this offset represents a systematic error in the arrival time of the original model. There are two drawbacks to this method of accounting for temporal offsets in the model: first, it can only account for a constant temporal offset $\Delta t$, rather than a distribution of uncertainties or a time-varying offset; second, this metric conflates the temporal alignment of the time series with the magnitudes of their predicted values, and thus does not necessarily characterize the model lag/lead time alone. Nonetheless, constant temporal offsetting is frequently used as a method to simply and quickly estimate model uncertainties, and as such remains available in MMESH.

\subsubsection{Dynamic Time Warping}\label{sec:MMESH:Performance:DTW}

The performance of a solar wind propagation model can be decomposed into two components: the performance in modeling the arrival time and the performance in modeling the magnitude of the solar wind time series. These two are essentially represented by the abcissa (i.e. the `independent' variable of time) and ordinate (i.e. the `dependent' variable, such as $u_{mag}$) pairs of a propagated time series, respectively. Theoretically, differences between the propagation model and data time series should be decomposable by first optimizing the alignment of the model relative to the data to characterize the performance in arrival time, then secondly measuring the residuals between the aligned model and data time series to characterize the performance in magnitude. Aligning the model to the data in this way is often done by manually identifying patterns of shocks in both time series and calculating the difference in their observation times \cite<e.g.>{Tao2005}. 

In practice, characterizing model performance in arrival time alone is not so straightforward, as the identification of patterns of shocks and shock-like structures in the solar wind data is often subjective. To more objectively define such structures, here we have ``binarized'' both the in-situ and propagation model time series data to identify extrema in both. The binarization process developed here involves taking the standard score (z-score) of the time derivative of a boxcar-smoothed time series and threshholding the result at a given significance level. This process has the end effect of identifying and isolating steep gradients in the time series of a given parameter, as would be expected in a shock, and is described in more detail in \ref{sec:Appendix:Binarization} and illustrated in \ref{fig:Binarized_Timeseries}. The binarization process was applied to the solar wind flow speed $u_{mag}$ time series in both the model-propagated and in-situ data sets. The boxcar-smoothing-widths used for each time series and in each epoch were found dynamically and are listed in Table \ref{tab:binarization_width}. Here a constant significance level of $3\sigma$, measured across the full duration of each time series, has then been used for binarization.

\begin{figure}[h]
    \centering
    \includegraphics[width=\textwidth]{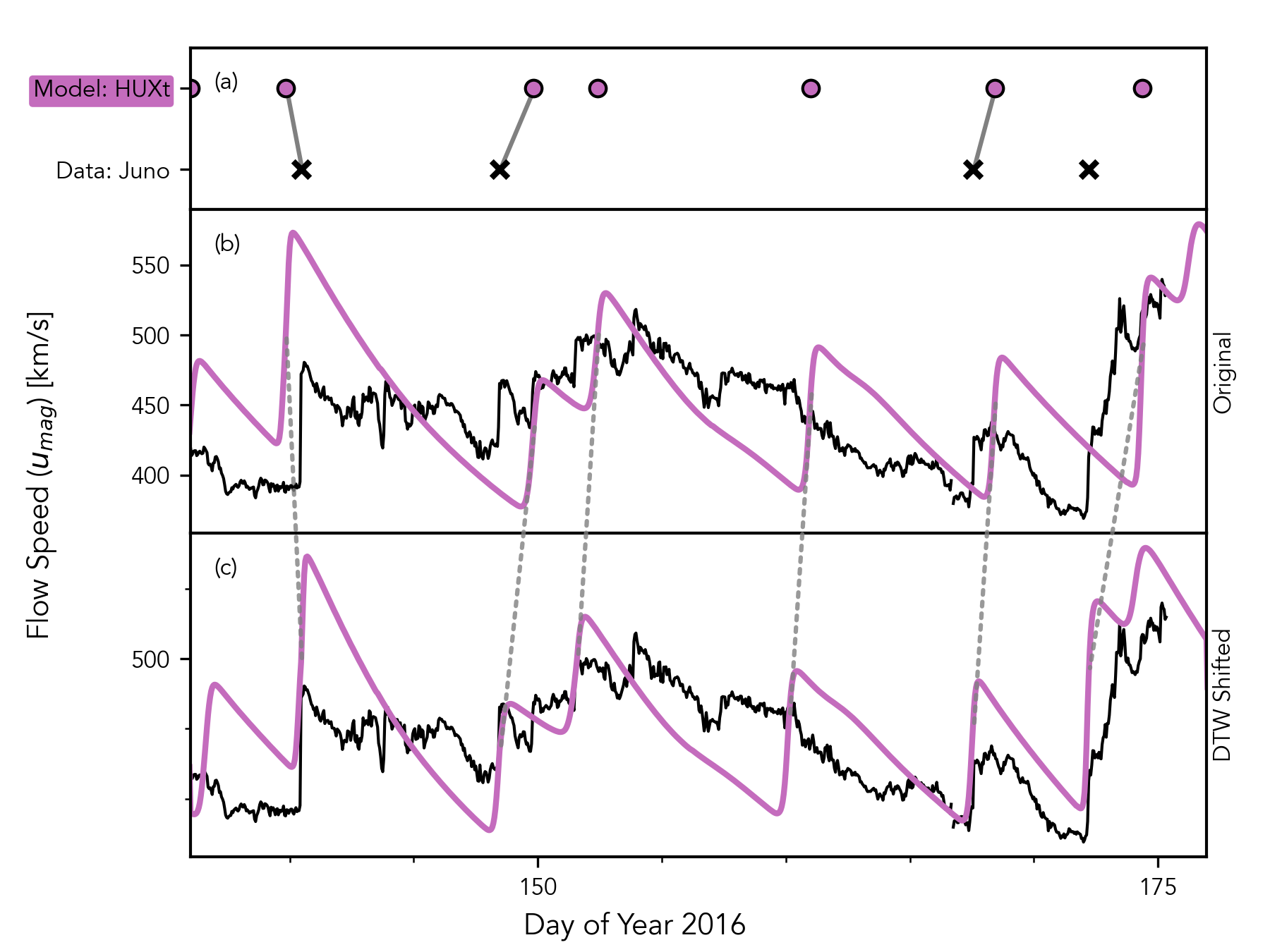}
    \caption{A composite diagram offering an overview of the dynamic time warping (DTW) process used to characterize model arrival time uncertainties. The (a) binarized model and data are shown as points representing the calculated extrema, with lines connecting model and data features which were identified to map to one another in the DTW process. The (b) original model and data time series are plotted to show the original alignment and may be compared to the (c) alignment of the warped model to the unchanged data, which demonstrates significantly reduced arrival time uncertainties. Dashed lines (b-c) connect the extrema identified in the original model to the same in the warped model; the horizontal component of these lines represents the offsets $\delta t$ used to warp the model to best match the data. These $\delta t$ are then taken as the distribution of arrival time uncertainties for the model. The temporal span here is 2016/05/15 - 2016/06/29, corresponding to the in-situ \textit{Juno} data shown in Table \ref{tab:Spacecraft}.}
    \label{fig:DTW_Overview}
\end{figure}

Identifying shocks and shock-like structures in the now-binarized time series is trivial; aligning the patterns of structures found in the model and data time series is not, and remains subjective if performed manually. For reproducability, here we employ an objective, automated method of aligning the two binarized time series based on the class of algorithms collectively known as dynamic time warping (DTW) \cite{Sakoe1978}. Qualitatively, the aim of DTW is to locally shift, stretch, and compress one time series to better resemble another; various adjustments can be made to the basic implementation of DTW in order to better match one time series to another, or to restrict matches to those which are physically realistic \cite[and references therein]{Keogh2001}. DTW has only recently been applied to space weather modeling problems; the calculated net distance has been suggested as a useful, multi-scale metric for measuring the performance of solar wind models by \citeA{Samara2022}, and the resulting alignments have been used to create more accurate boundary conditions for solar wind propagation models by \citeA{Owens2021}. Within MMESH, the \texttt{dtw-python} package for the Python programming language developed by \citeA{Giorgino2009} is employed to warp the modeled time series to more closely resemble the in-situ data. The recommended usage, and that which will be followed in this discussion, is to use DTW to align the binarized model solar wind flow speed to the binarized measured flow speed, as the flow speed generally shows the clearest signatures of shocks and shock-like structures after binarization. Both the binarization and DTW methods within MMESH can, however, be applied independently to any of the propagated solar wind quantities (i.e. $n$, $u_{mag}$, $p_{dyn}$, or $B_{IMF}$) at the discretion of the user.

An overview of the two-series implementation of DTW is illustrated in Figure \ref{fig:DTW_Overview} and described here. This approach involves calculating the Euclidean distance between every permutation of the elements of each series, resulting in a two-dimensional matrix; a path, or alignment curve, through this matrix is then computed which minimizes the net distance, and this serves to effectively align the input modeled time series to best match the data by reindexing the former. DTW is here applied to the binarized time series data (Figure \ref{fig:DTW_Overview}a) in order to eliminate the effect that each series' amplitude may have on the alignment calculation. From the aligned time series, tie points connecting the model time series to the data are then chosen from the alignment curve for each matching pair of model-data extrema (Figure \ref{fig:DTW_Overview}a). The original model time series (Figure \ref{fig:DTW_Overview}b, in purple) is then warped according to an interpolation of these tie points, which represents both the offsets of the matched extrema and the linear interpolations at each abcissa between these. Other forms of interpolation may be used to find warping values between the binarized tie points; a linear interpolation is used here for simplicity. The result is a warped time series which is better aligned with the spacecraft data (Figure \ref{fig:DTW_Overview}c). While this process uses the binarized solar wind flow speed to compute the alignment, every parameter within a given model can then have the same warping applied to it. This allows for better alignment between all parameters, not just $u_{mag}$, by implicitly assuming that the input model parameters are aligned correctly with one another, and misaligned only relative to the measured data.

\begin{figure}
  \begin{center}
    \includegraphics[width=0.48\textwidth]{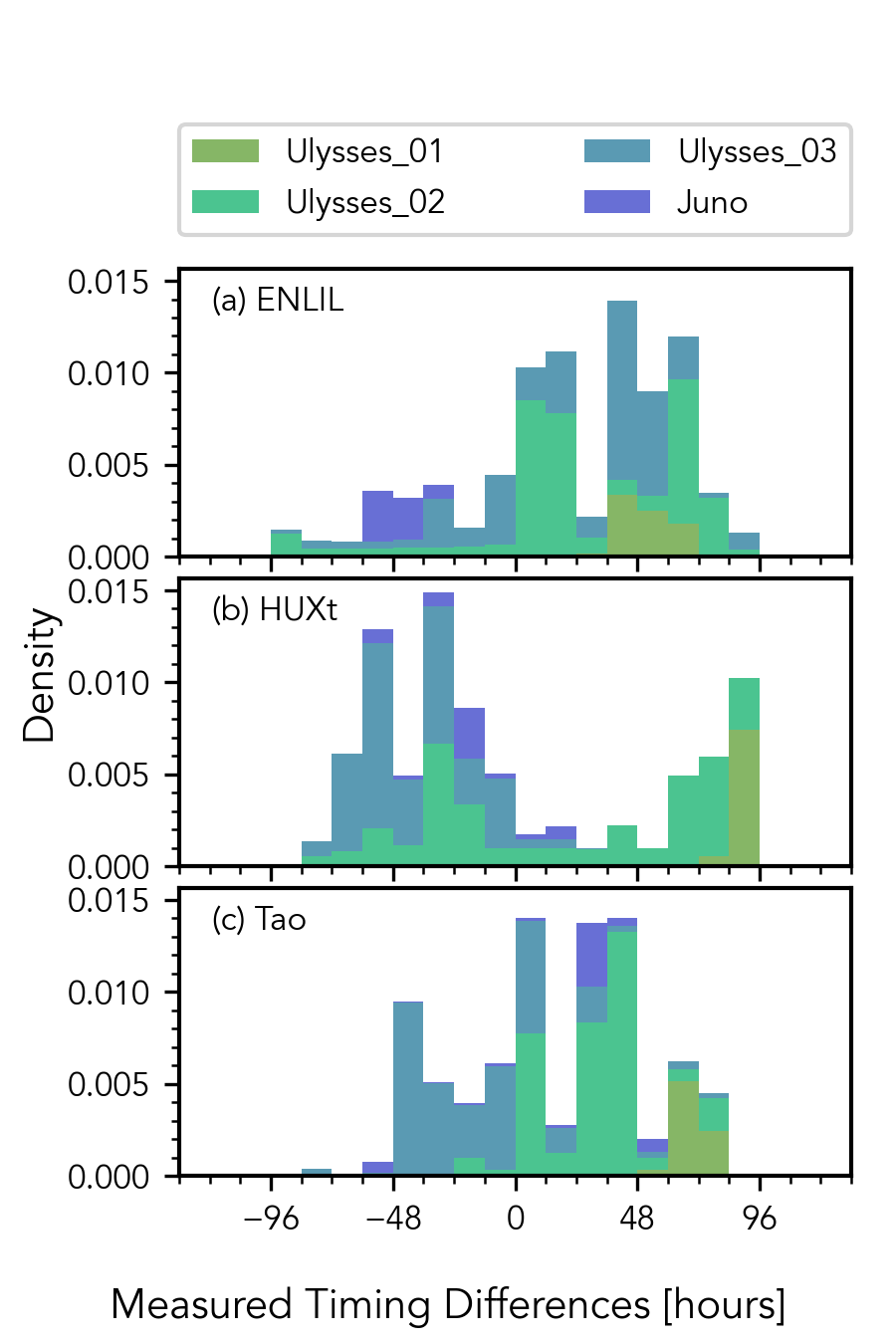}
  \end{center}
  \caption{Histograms showing the distribution of total temporal shifts needed to best align each model with the in-situ data for the all spacecraft epochs, as found using dynamic time warping (DTW) as described in the text. The means of these distributions are equivalent to intrinsic shock arrival time errors, or timing biases, and the widths are representative of timing uncertainties.}
  \label{fig:DTW_Widths}
\end{figure}

For this demonstration of MMESH, DTW was used to align the binarized solar wind plasma flow speeds from each model to that of the data in each spacecraft epoch. Limits were placed on the DTW algorithm to ensure the resulting warped time series was physically meaningful: the maximum offsets allowed were $\pm4$ days ($\pm96$ hours), chosen to be representative of the maximum temporal offsets measured in other studies \cite{Tao2005, Zieger2008}. 
The first value of the modeled time series is forced to align to the first value of the measured time series by the DTW algorithm used here, as is the final value of the modeled time series to that of the measured. To account for this, the DTW process was applied to the same $2n+1$ models with constant temporal offsets in the range $[-n, n]$ and with step size $\Delta t$ as was previously discussed in Section \ref{sec:MMESH:Performance:Constant}. The optimal alignment within these $2n+1$ DTW results was found by maximizing the correlation coefficient of the warped model plasma flow speed $u_{mag}$ to the data divided by the quasi-$1\sigma$ half-width of the distribution of total temporal offsets $r/\sigma_{P}$ (i.e., both constant and dynamic temporal offsets combined).
The total distributions of temporal offsets in each model are illustrated in Figure \ref{fig:DTW_Widths} for reference. These distributions are not normally distributed, suggesting that the uncertainties in the modeled solar wind arrival times are not random, and are not centered at zero, indicating biases in the modeled arrival times.

\begin{figure}[h]
    \centering
    \includegraphics[width=0.5\textwidth]{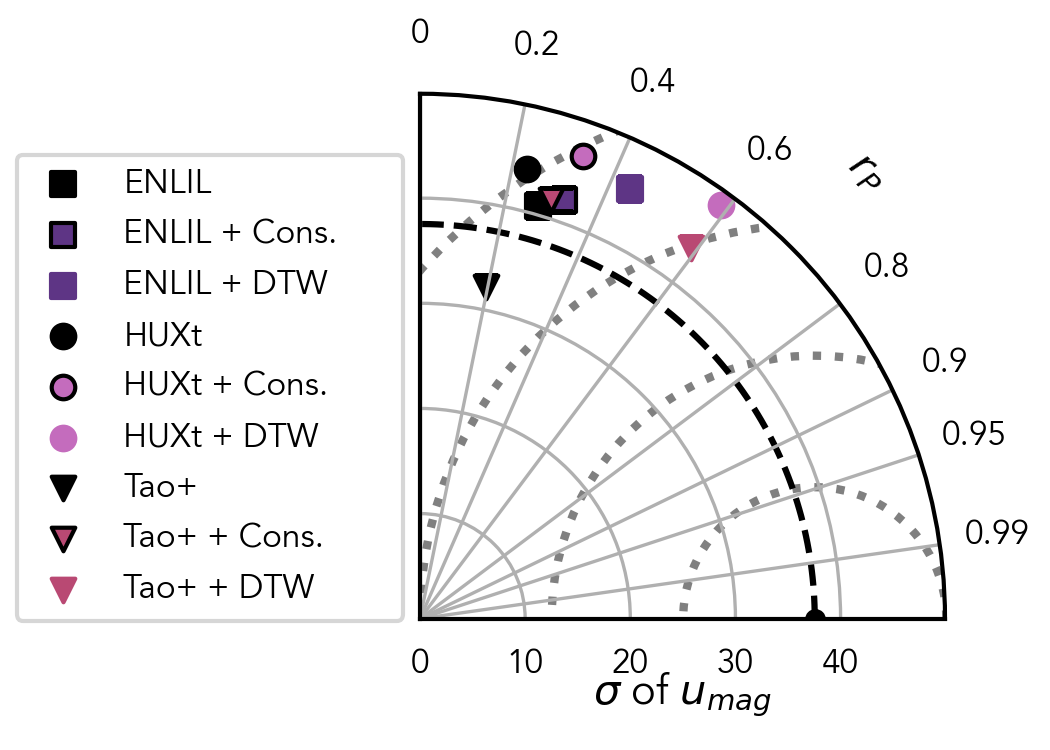}
    \caption{A Taylor Diagram showing the performance of each model, before and after temporal shifting, relative to the in-situ \textit{Juno} solar wind data. The unshifted models (black symbols) all have correlation coefficient $r$ in the range $0.2{\sim}0.3$. Both constant time offsetting (outlined symbols) and DTW (full color symbols) improve the correlation coefficients of all models, but DTW improves the correlation coefficient more ($r$ between $0.3{\sim}0.4$ compared to $r$ between $0.4{\sim}0.6$, respectively). Employing time-varying temporal shifts is beneficial to matching the models to the data more closely.}
    \label{fig:TD_Input}
\end{figure}

\subsection{Prediction of Time-Varying Model Timing Uncertainties}\label{sec:MMESH:Prediction}
The cross-model consistent characterization of systematic timing biases and uncertainties already discussed allows the performance of the solar wind models to be quantitatively compared to one another. As the methods discussed in Section {\ref{sec:MMESH:Performance}} rely on direct comparison to contemporaneous data, however, the timing biases and uncertainties cannot be empirically quantified in the absence of in-situ data-- the main use case for solar wind modeling. To circumvent this, the distribution of timing uncertainties, as illustrated in Figure \ref{fig:DTW_Widths}, could be considered invariant in time and propagated as such; this method of propagating timing uncertainties is supported by MMESH. As these timing uncertainties and biases are known to vary in time, as can be seen by the different spacecraft epochs covered in Figure {\ref{fig:DTW_Widths}}, this method has the drawback of explicitly overestimating the uncertainties at any given time.

Alternatively, MMESH also supports a simple- or multiple-linear regression model description of the timing uncertainties. 
Multiple linear regression models are simple models which describe one continuous target variable as a linear combination of multiple continuous predictor variables; simple linear regression refers to the special case of a single predictor variable. The coefficients calculated for each predictor variable thus describe the contributions of each to the target variable. Similarly, the estimated standard deviation on these coefficients gives a sense of the relative importance of each predictor: relative to the coefficient value, a large standard deviation denotes a less significant predictor, with the opposite being true for a relatively small standard deviation. The linear regression method thus allows the propagation model timing uncertainties to be estimated even in the absence of in-situ data for comparison, provided the values of each predictor variable are known.

\begin{figure}
    \centering
    \includegraphics[width=\textwidth]{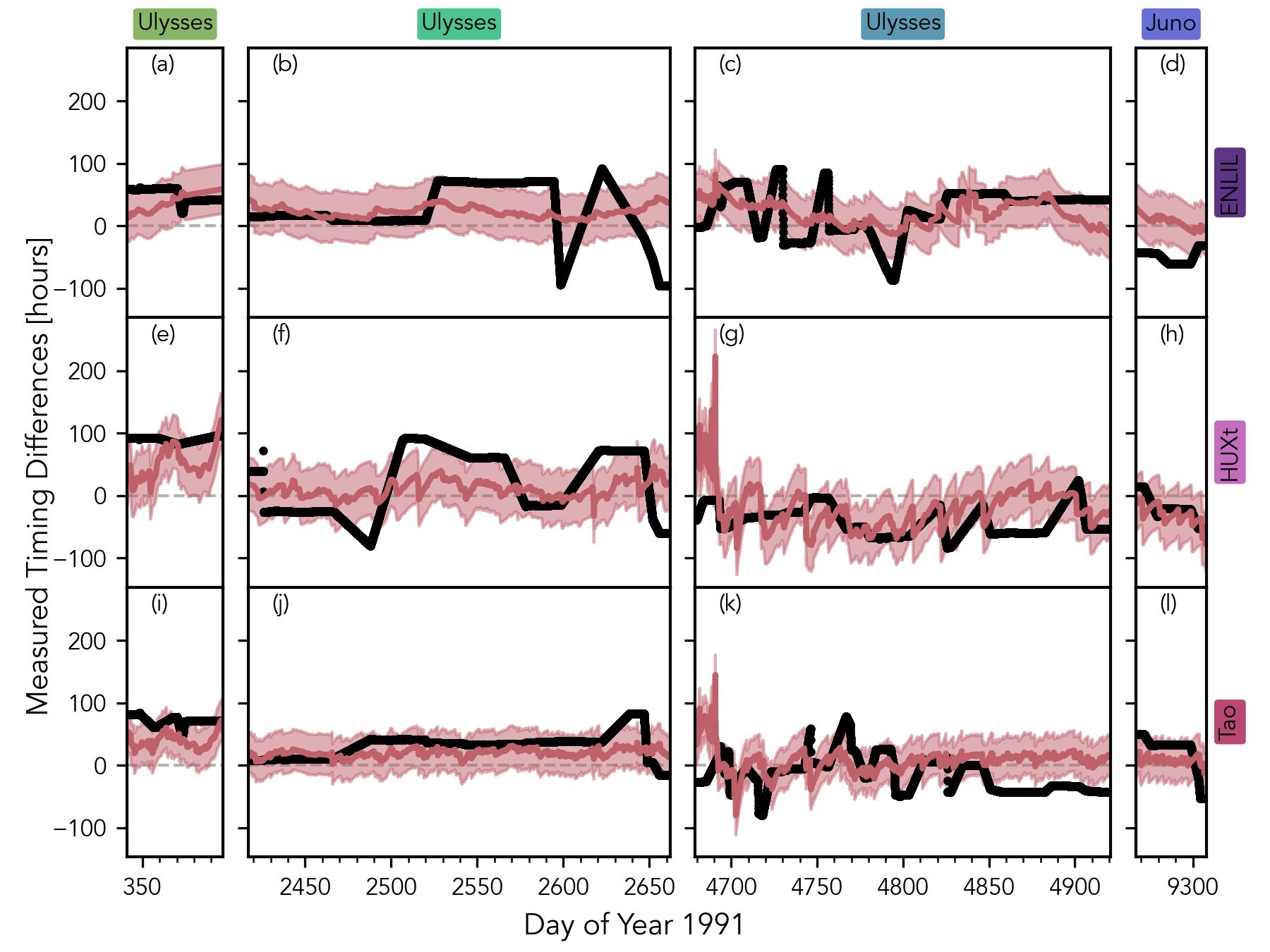}
    \caption{Plots of the measured temporal offsets (black lines) from DTW for each model-spacecraft-epoch set (e.g., a-d for ENLIL, e-h for HUXt, and i-l for Tao+), along with the multiple linear regression (MLR) fit to the temporal offsets found by fitting the offset time series with the parameters described in the text (red lines). The time spans of each column correspond to the spans listed for each spacecraft epoch in Table {\ref{tab:Spacecraft}}. While the independent parameters add significant variation in time, they nonetheless describe the emprirical timing uncertainties and systematics fairly well. The $1\sigma$ prediction uncertainities in the MLR fit (shaded red regions) are also plotted.}
    \label{fig:MLR_Example}
\end{figure}

As the input propagation models do not propagate measurement error, the arrival time uncertainties characterized previously are present due to the limitations of these MHD-based models, each of which makes different simplifications of the physics describing the solar wind. These simplification give rise to correlations between the timing uncertainties in these models and other physical parameters describing the solar wind environment.
Timing uncertainties in models with an inner boundary set by near-Earth measurements often trend with target-Sun-observer (TSO) angle in heliolongitude, in at least magnitude if not also in sign \cite{Tao2005, Zieger2008}. Physically, this trend represents increasing uncertainty in the solar wind conditions as separation in heliolongitude (or Carrington longitude) increases away from the measurement point. While less commonly used, the offsets are expected to trend with the TSO angle in heliolatitude in a similar way, as the solar wind flow speed is known to be strongly ordered in heliolatitude during solar minimum \cite{McComas2003, McComas2008}. This well-ordered structuring with heliolatitude breaks down during solar maximum \cite{McComas2003}, which further suggests a physical connection between the offsets and the 11-year solar cycle. A final reasonable expectation is that the timing systematics and uncertainties are related to the models solar wind flow speed $u_{mag}$. This comes from the assumption that the propagation model is more likely to lag the data when underestimating the solar wind flow speed and more likely to lead when overestimating; if the underestimates tend to have lower magnitudes and overestimates tend have larger magnitudes, then a trend between modeled solar wind flow speed and temporal offset is expected.

These physical relationships between total, time-variable model offsets and descriptive parameters about the state of the solar wind can be leveraged to estimate the model offsets in the absence of simultaneous in-situ data. Here, multiple linear regression has been employed to use all of these physical parameters (i.e. TSO angle in heliolongitude and heliolatitude, solar cycle phase, and modeled $u_{mag}$) as predictors of the time-variable timing uncertainties and biases by fitting the predictors to the combined spacecraft epochs during which simultaneous in-situ measurements are available, as illustrated in Figure \ref{fig:MLR_Example}. Despite its simplicity, the multiple linear regression technique matches the known temporal offsets well. The combination of parameters used here accounts for $12\%$ of the variation in the measured timings for the ENLIL model (i.e., $R^2=0.12$), $37\%$ in the HUXt model, and $20\%$ in the Tao+ model.

\subsection{Multi-Model Ensemble}\label{sec:MMESH:MME}
An MME is now created by the combination of the propagation models.
MMESH supports the creation of MMEs from input propagation models alone, from propagation models with characterized timing uncertainties, whether through constant time offsetting or dynamic time warping, and from propagation models de-trended (i.e. warped) to account for timing biases with propagated uncertainties. Here, this final type of MME is created from the ENLIL, HUXt, and Tao+ solar wind propagation models warped according to the timing biases estimated by multiple linear regression to the multi-epoch in-situ dataset, and timing uncertainties propagated through.

For simplicity, an equal weights average of the each input model is taken. While there is some evidence that carefully-chosen weighting schemes may improve model performance \cite{Guerra2020}, more complicated weighting schemes may also decrease model performance compared to the equal weights, making equal weighting the more robust choice \cite{Genre2013}. Thus, the only improvement on the simple equal-weights averaging scheme we impose is to set the weight to $0$ when a model does not yield an output at a given time step, whether due to the model's design (e.g. the lack of parameters other than solar wind flow speed in HUXt) or a lack of access to more recent models. The resulting MME of solar wind flow speed is shown in Figure \ref{fig:Ensemble_Comparison}a, superimposed on the in-situ measurements of the \textit{Juno} spacecraft during the missions's cruise phase (cf. Figure \ref{fig:ModelVelocities}).

\subsubsection{Model Performance}\label{sec:MMESH:MMEPerformance}

\begin{figure}
    \centering
    \includegraphics[width=\textwidth]{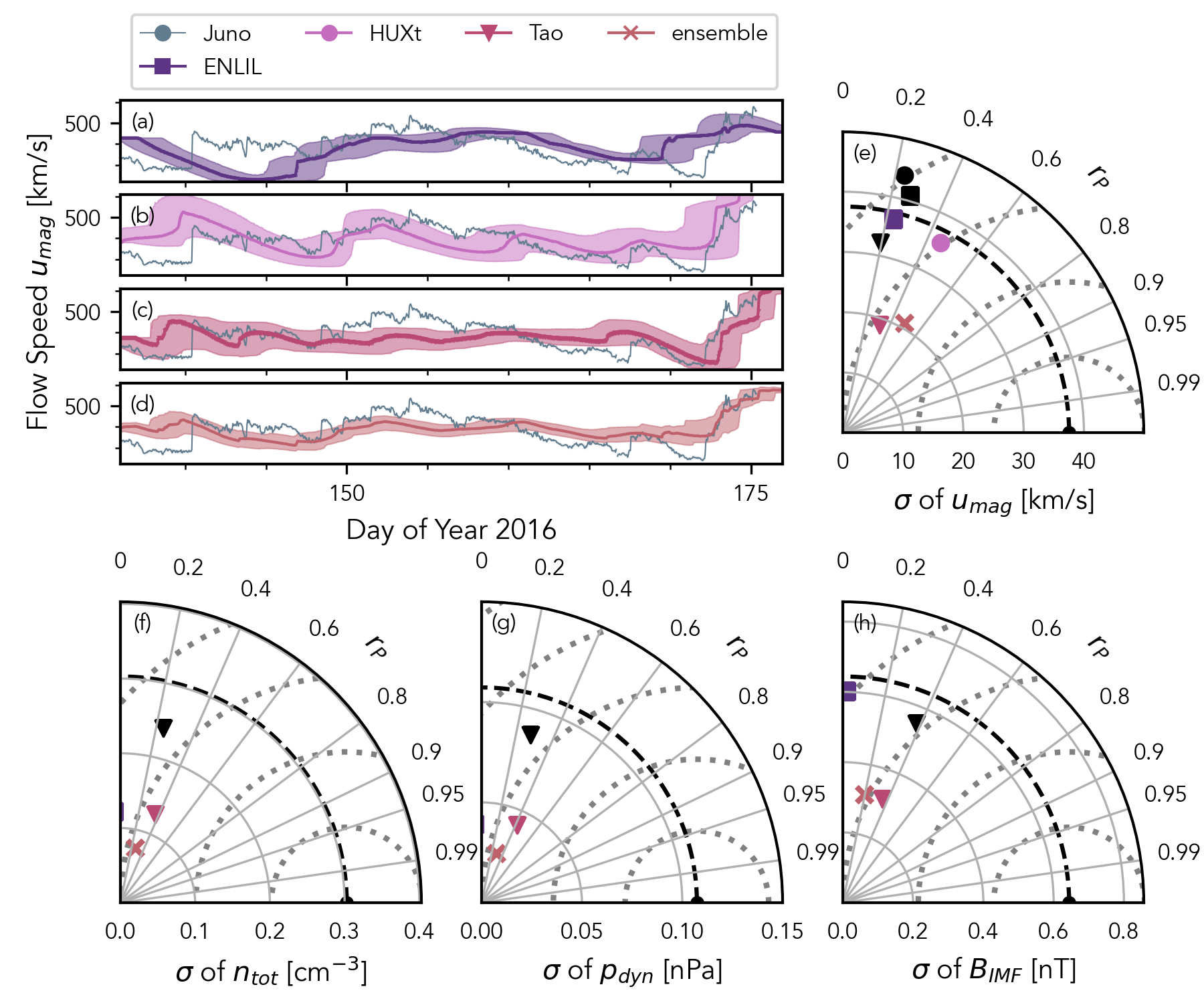}
    \caption{The solar wind flow speed $u_{mag}$, with timing uncertainties characterized by DTW and MLR applied over all spacecraft epochs, for (a) ENLIL, (b) HUXt, (c) Tao, and (d) the MME, compared to in-situ \textit{Juno} data in each. The performance of the MME is summarized in the (e) Taylor diagram for $u_{mag}$, which illustrates that the MME outperforms all input models for this parameter; the Taylor diagram includes both the multi-epoch MLR-adjusted input models (colored symbols) and the original input models (black symbols) for comparison. Additional Taylor diagrams for (f) the total solar wind density $n_{tot}$, (g) the solar wind dynamic pressure $p_{dyn}$, and (h) the IMF magnitude $B_{IMF}$ are included to show the performance of the MME in these parameters. As HUXt does not contribute to these parameters, the MME slightly underperforms Tao+.}
    \label{fig:Ensemble_Comparison}
\end{figure}

 The performance of this multi-model, multi-epoch ensemble is summarized in Figure \ref{fig:Ensemble_Comparison}. All of the model time series, including that of the ensemble, show decreased standard deviations in Figures \ref{fig:Ensemble_Comparison}a-d. This results from considering the distribution of timing uncertainties in calculating the mean values for each time series. Due to the wide distributions of measured timing uncertainties for each model, the shifted fore-shocks in the solar wind appear more `smoothed out' at all times when compared to Figures \ref{fig:ModelVelocities}a-c. 
 
 Nonetheless, the prediction efficiency of the MME in nowcasting solar wind flow speed $u_{mag}$ is improved compared to any individual input model, either before or after DTW- and MLR- based shifting during the \textit{Juno} cruise epoch (Figure \ref{fig:Ensemble_Comparison}e). The predicted flow speed $u_{mag}$ of the MME ($r=0.49$) outperforms ENLIL by $110\%$ ($r=0.23$), HUXt by $7\%$ ($r=0.46$), and Tao+ by $51\%$ ($r=0.32$) in correlation coefficient and achieves a centered root-mean-square difference (RMSD$=32.8$) $28\%$ lower than ENLIL (RMSD$=45.9$), $14\%$ lower than HUXt (RMSD$=38.1$), and $9.1\%$ lower than Tao+ (RMSD$=36.1$). As HUXt does not contribute to parameters in the MME other than $u_{mag}$, and the performance of ENLIL beyond $u_{mag}$ is poor here (i.e., ENLIL is anticorrelated with the data in Figure \ref{fig:Ensemble_Comparison}f-h, as evidenced by its absence from the Taylor diagrams), the MME underperforms Tao+ in $n_{tot}$, $p_{dyn}$, and $B_{IMF}$ by $12\%-24\%$ in correlation coefficient with $5\%-8\%$ larger RMSD. These shortcomings of the MME are thus slight, and would likely be reduced further or eliminated by: analyzing epochs where ENLIL performs more similarly to Tao+; adding new solar wind propagation models to the MME discussed here; or by changing the model weighting scheme to reflect overall model performance. Additionally, it is worth noting that, after being shifted within the MMESH framework, the Tao+ model outperforms the original, unshifted model in correlation coefficient by $55\%$ in solar wind density $n_{tot}$ and $48\%$ in solar wind dynamic pressure $p_{dyn}$, while underperforming by $6\%$ against the original in predicts IMF magnitude $B_{IMF}$. The MMESH-based uncertainty estimation and de-trending thus generally results in an improved solar wind model even in the single-model case.
 
As the MME depends directly on the performance of the input models, improvements to these input models would results in an improved MME. For instance, using data assimilation to take data from multiple solar wind monitors has been shown to reduce uncertainties in the resulting models by breaking the degeneracy between time and solar longitude present in inner boundaries derived from single spacecraft {\cite<e.g.>{Barnard2023}}. Advances in the physical understanding of the solar wind may also allow improved models to be developed. As an example, recent results from the Parker Solar Probe (PSP) suggest that the acceleration of the solar wind near the sun is more impulsive than previously expected {\cite{Bunting2024}}. Implementation of solar wind acceleration profiles which more closely match these new results may improve model timing uncertainties.

\section{\textit{Juno}-epoch Solar Wind MME for Jupiter}\label{sec:JunoEpochResults}
\begin{figure}[b]
    \centering
    \includegraphics[width=\textwidth]{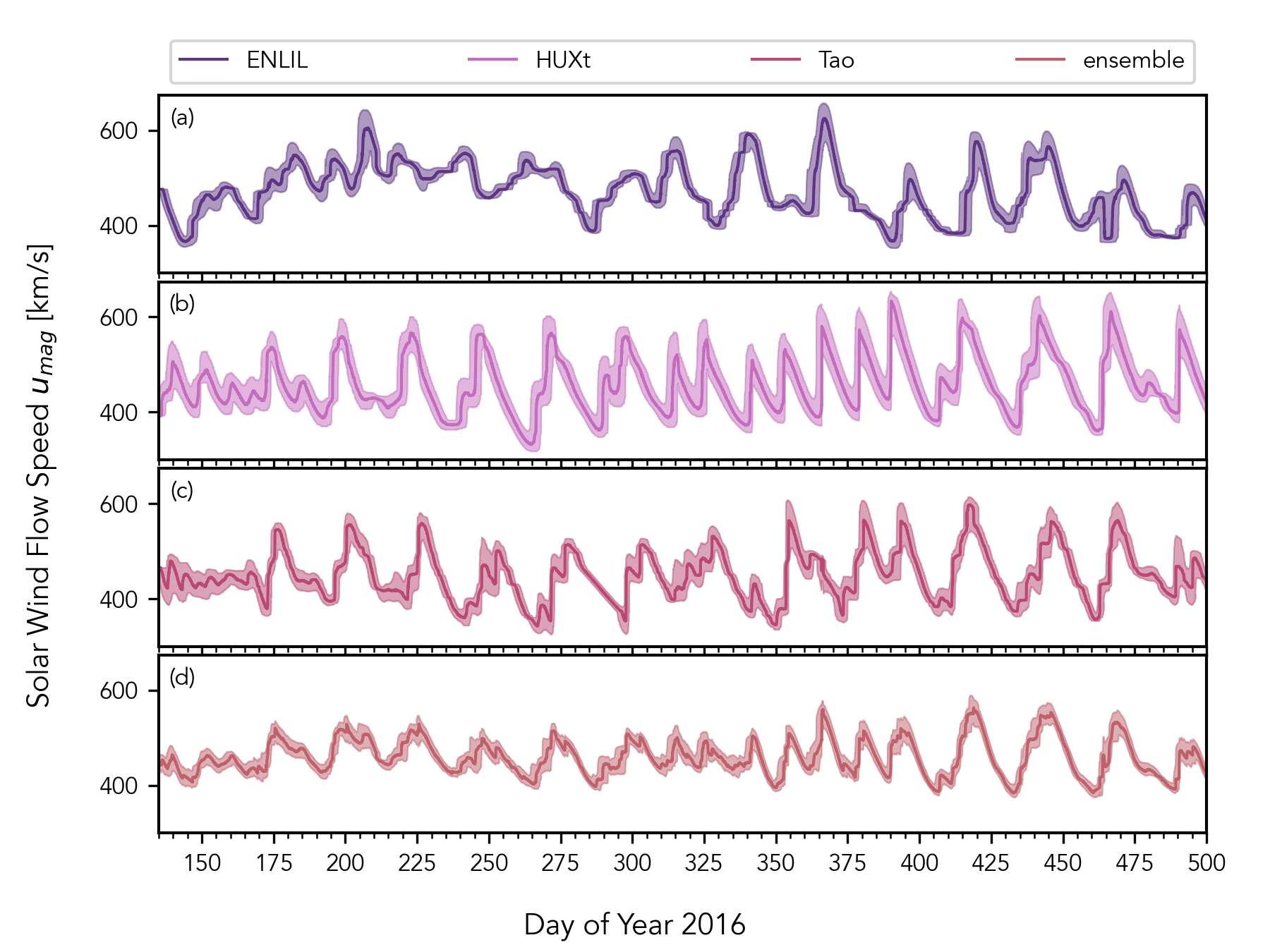}
    \caption{A $12$-month subset of the \textit{Juno}-era solar wind flow speed $u_{mag}$ results, adjusted for timing biases measured using DTW and characterized using MLR, for the (a) ENLIL, (b) HUXt, (c) Tao+, and (d) MME, presented here starting during \textit{Juno}'s approach to Jupiter in May 2016. The $1\sigma$ uncertainties in the solar wind flow speed $u_{mag}$ are shown in each panel (shaded regions). Based on the results discussed here, the MME is expected to significantly outperform each of the component models in predicting the solar wind flow speed $u_{mag}$.}
    \label{fig:Timeseries_Ensemble}
\end{figure}

Now that the multi-model epoch system has been fully described, all that remains is to generate MMESH-propagated solar wind for a useful epoch. Here we have chosen to run the ensemble for Jupiter contemporaneously with the \textit{Juno} mission, beginning with the orbital insertion of the spacecraft (2016/07/04) and continuing seven years through mid-2023 (2023/07/04), in order to provide valuable context for the upstream conditions near Jupiter during \textit{Juno}'s mission. A subset of the ensemble model results are shown in Figure \ref{fig:Timeseries_Ensemble}, along with the results of the component models, spanning the first year of coverage provided by this MME.
The results of this specific MME are available at \url{https://doi.org/10.5281/zenodo.10687651} \cite{Rutala2024_JupiterMME_v1.0.0}; more generally, the results of this Jupiter MME along with any future updates to improve its predictive power or extend the temporally coverage will be available, and documented, at \url{https://doi.org/10.5281/zenodo.10687650} \cite{Rutala2024_JupiterMME}.

\section{Summary and Conclusions}\label{sec:Summary}
Here we have introduced MMESH, a Multi-Model Ensemble System for the Heliosphere, and described one use-case of this system to create a multi-model ensemble of the outer heliosphere solar wind near Jupiter through the first $7$ years of \textit{Juno} mission, spanning 2016/07/04 -- 2023/07/04.

MMESH provides a framework with two central objectives: first, to allow easy characterization of solar wind propagation model performance; and second, to create multi-model ensembles of the solar wind. The first objective is crucial to statistically evaluating the strengths of the various solar wind propagation models available, as the original discussions of the performance of these models often quote different statistics or span non-overlapping epochs of the solar wind and thus cannot be compared one-to-one. Further, characterization of model performance yields an estimate of the model uncertainty, a quantity which is not provided internally by any model discussed here but which is essential for statistical analyses. With the second objective, we aim to create reliable composite models of the solar wind by combining physics-based solar wind propagation models with their estimated variances to be used in statistical analyses of solar-wind-magnetosphere interactions throughout the solar system. The strength of ensemble modeling lies in leveraging the different strengths of the constituent models, and so these two objectives are closely intertwined.

MMESH additionally includes a method to compare biases and variances in the model timing to physical parameters across disparate epochs prior to creating an ensemble. The objective of this multi-epoch method is to de-trend biases in the model timing which may arise from the various assumptions and simplifications made by each model. De-trending is performed here through multiple linear regression (MLR) of the measured model timing biases with a subset of the physically reasonable parameters with which model performance is expected to vary. The phase of the solar cycle, difference in heliolongitude and heliolatitude between the model target and the observer, and the modeled solar wind flow speed are all reasonable and considered here. As estimation of the model timing biases and variances is only possible when contemporaneous in-situ data are available for comparison, the spans over which the MLR de-trending can be performed are limited. The MLR de-trending is made more robust by considering multiple disparate epochs during which spacecraft data are available.

Using all of these methods, a multi-model ensemble of the solar wind conditions at Jupiter during the \textit{Juno}-epoch (2016/07/04 - 2023/07/04) has been created by combining three physics-based solar wind propagation models (ENLIL, HUXt, and Tao+). The version of this multi-model ensemble discussed here is available at \url{https://doi.org/10.5281/zenodo.10687651} \cite{Rutala2024_JupiterMME_v1.0.0} and the latest release of is available at \url{https://doi.org/10.5281/zenodo.10687650} \cite{Rutala2024_JupiterMME}; the MMESH code used to generate the multi-model ensemble is available via \url{https://github.com/mjrutala/MMESH} \cite{Rutala2024_MMESH_Code}. Biases and variances in each model's timing were characterized for four epochs during which \textit{Ulysses} or \textit{Juno} data were available for comparison, spanning in total from 1991/12/08 -- 2016/06/29. The model timing biases were then de-trended using MLR to the heliolatitude and modeled flow speed, which were determined to provide the best balance between describing the timing biases and overfitting. The biases in the three constituent solar wind models were corrected according to the MLR equation for the full ensemble span of 2016/07/04 -- 2023/07/04 and combined. The resulting ensemble model outperforms all of the constituent models relative to the \textit{Juno} cruise data immediately preceding this epoch; looking at solar wind flow speed $u_{mag}$, the ensemble has a correlation coefficient of 0.49 ($110\%$ increase over ENLIL, $7\%$ increase over HUXt, and $51\%$ increase over Tao+, after accounting for timing offsets in each). The improved upstream solar wind monitoring capabilities demonstrated by this ensemble are available to be downloaded and used immediately, and should prove crucial to ongoing and future in-situ studies of the Jovian magnetosphere using {\textit{Galileo}}, {\textit{Juno}}, JUICE, and {\textit{Europa Clipper}}, as well as remote sensing studies using observatories such as JWST, HST, and \textit{Chandra}.


%
%
%
%
\appendix

\section{Time Series Binarization} \label{sec:Appendix:Binarization}

\begin{figure}
    \centering
    \includegraphics[width=\textwidth]{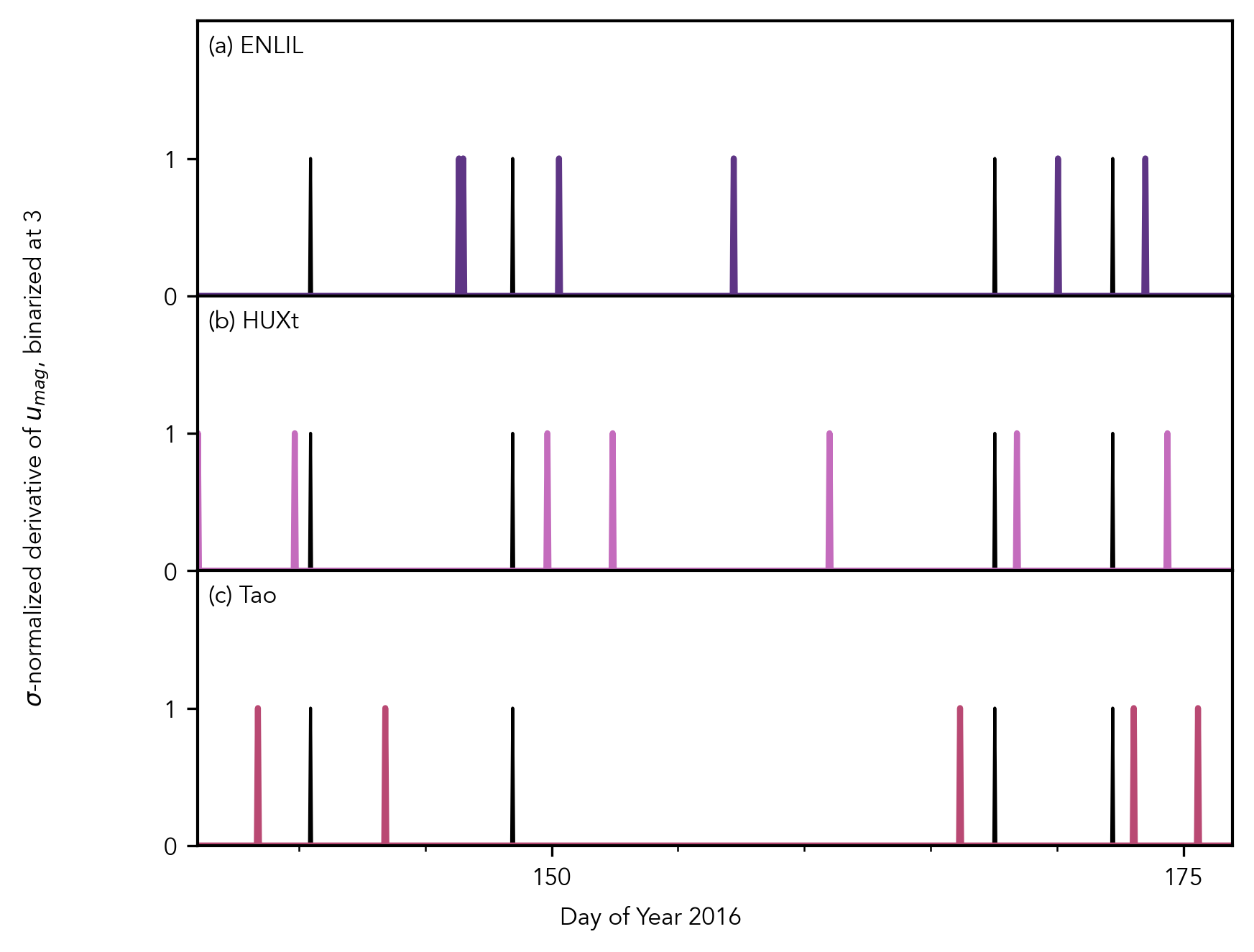}
    \caption{Binarized time series of the solar wind flow speed $u_{mag}$ for the (a) ENLIL, (b) HUXt, and (c) Tao+ solar wind propagation models, with the binarized time series of the in-situ \textit{Juno} data superimposed on each (black lines). The time-derivatives of all these series have been binarized at a value of $3\sigma$, such that each `spike' represents a change in the time-derivative of $3\sigma$ or larger.}
    \label{fig:Binarized_Timeseries}
\end{figure}

Here, the measured and modeled magnitude of the solar wind flow speed $u_{mag}$ is post-processed by first smoothing the series, then taking the standard score of its time derivative. Smoothing is accomplished by taking a rolling boxcar average of the flow speed $u_{mag}$. Smoothing in this way serves as a low-pass filter, allowing the recovery of the large-scale shape of the time series while ignoring small-scale fluctuations, which may dominate in in-situ spacecraft measurements. The time derivative of the flow speed time series $u_{mag}(t)$ is chosen in order to better identify the transition of a spacecraft or model trajectory through a slow-fast wind interface; these increases in solar wind flow speed occur over timescales less than $1 \mathrm{hour}$ and are more easily identifiable than changes in other solar wind parameters, which typically occur over longer timescales. The standard-score of the time series, or the time series normalized to its own standard deviation, allows for direct comparison of the relative changes between different time series which may have widely varying mean values.

Binarization requires subjective input of a boxcar-smoothing-width and significance level for each time series, however these parameters are partially degenerate with one another-- a smaller smoothing window and a higher significance level will yield similar results to a larger window with lower significance level. To limit subjectivity, boxcar-smoothing-widths are found for each time series within a given epoch as the smallest width which, when applied to each time series before the derivative is taken, results in an equal series standard deviation to the smallest such standard deviation in the epoch. Qualitatively, this is the boxcar-smoothing-width required to make each time series look as `smooth' as the `smoothest' time series of the epoch. The boxcar-smoothing-widths used for each time series and in each epoch are listed in Table \ref{tab:binarization_width}.

\begin{table}[!ht]
\caption{Boxcar-smoothing widths for binarization, in hours}
\centering
\begin{tabular}{c c c c c}
\hline
Source & \multicolumn{4}{c}{Epoch}\\
 & Ulysses 01 & Ulysses 02 & Ulysses 03 & Juno \\
\hline
    in-situ & 5 & 15 & 2 & 7 \\
    ENLIL  & 8 & 1 & 4 & 5 \\
    HUXt & 1 & 9 & 1 & 1 \\
    Tao+  & 6 & 11 & 2 & 10 \\
\hline
\label{tab:binarization_width}
\end{tabular}
\end{table}

%
%

%

%

\section*{Open Research}

Pre-processed spacecraft data were acquired from the Goddard Space Flight Center Space Physics Data Facility (SPDF) service \cite{Ulysses_SPDF_DataSet}, except for the \textit{Juno} in-situ data, which were instead acquired from \citeA{Wilson2018} (plasma data) and \citeA{Juno_Cruise_AMDA_FGM} (magnetic field data). Context for these data in the form of F10.7 Solar Radio Flux was obtained from the Solar Radio Monitoring Program (\url{https://www.spaceweather.gc.ca/forecast-prevision/solar-solaire/solarflux/sx-en.php}) at the Dominion Radio Astrophysical Observatory (DRAO) with additional processing by the NOAA National Centers for Environmental Information (\url{https://www.ncei.noaa.gov/}). These data were accessed via the LASP Interactive Solar Irradiance Datacenter (LISIRD) (\url{https://lasp.colorado.edu/lisird/}). Ephemeris information was obtained by use of the NASA Navigation and Ancillary Information Facility (NAIF) SPICE toolkit (\url{https://naif.jpl.nasa.gov/naif/}). Simulation results for the ENLIL solar wind propagation model (version 2.8f) have been provided by the Community Coordinated Modeling Center (CCMC) at Goddard Space Flight Center through their publicly available simulation services (\url{https://ccmc.gsfc.nasa.gov}). The ENLIL Model was developed by D. Odstrcil at George Mason University \cite{Odstrcil2003}.

The MMESH code is available at \url{https://github.com/mjrutala/MMESH} \cite{Rutala2024_MMESH_Code}, and includes the routines used to create the figures shown here. The \textit{Juno}-epoch solar wind MME for Jupiter presented here is available via \url{https://doi.org/10.5281/zenodo.10687651} \cite{Rutala2024_JupiterMME_v1.0.0}, and future updates to this MME will be accessible via \url{https://doi.org/10.5281/zenodo.10687650} \cite{Rutala2024_JupiterMME}.






\acknowledgments
MJR's and CMJ's work at DIAS was supported by Science Foundation Ireland Grant 18/FRL/6199. MJO and LB are part-funded by Science and Technology Facilities Council (STFC) grant number ST/V000497/1. ARF's work was supported by Irish Research Council Government of Ireland Postdoctoral Fellowship GOIPD/2022/782. SAM is involved with the European Union Horizon Europe Project No. 101082164 (ARCAFF).



%
%
\bibliography{MMESH}



%
%
%
%
%

\end{document}